\documentclass[12pt,A4]{article}
\usepackage{amssymb,amsfonts,amsthm,amsmath}

\usepackage{epsf}
\usepackage{graphicx}
\usepackage{placeins}
\usepackage{setspace}

\def\hang{\hangindent\parindent}
 \def\rf{\par\noindent\hang}

\begin{document}

\baselineskip=21pt

\begin{center} \Large{{\bf
Large-Sample Confidence Intervals for the Treatment Difference
in a Two-Period Crossover Trial, Utilizing Prior Information}}
\end{center}

\bigskip

\begin{center}
\large{ {\bf Paul Kabaila$^*$ and Khageswor Giri}}
\end{center}

\begin{center}
{\sl Department of Mathematics and Statistics, La Trobe
University, Victoria 3086, Australia}
\end{center}

\medskip

\noindent {\bf Abstract}  
%{\bf (37 words, under the 40 word limit)}

\medskip

Consider a two-treatment, two-period crossover trial, with responses that are
continuous random variables. We find a large-sample frequentist $1-\alpha$ confidence interval
for the treatment difference that utilizes the uncertain prior information that there
is no differential carryover effect.

\bigskip
\medskip

\noindent {\it Keywords:} Differential carryover effect; Prior
information; Two-period crossover trial.

\vbox{\vskip 6.5cm}

%\bigskip

\noindent $^*$ Corresponding author. Address: Department of
Mathematics and Statistics, La Trobe University, Victoria 3086,
Australia; Tel.: +61-3-9479-2594; fax: +61-3-9479-2466.
{\it E-mail address:} P.Kabaila@latrobe.edu.au.

%\renewcommand{\baselinestretch}{1.5}

%\bigskip

\newpage

\noindent {\bf 1. Introduction}

\medskip

We consider a two-treatment two-period crossover trial, with responses that are continuous random
variables. This design is very popular in a wide range of medical and other applications,  see e.g. 
Jones and Kenward (1989) and Senn (2006).
The purpose of this trial is to carry out inference about the difference $\theta$ in the effects
of two treatments, labelled A and B. Subjects are randomly allocated to either group 1 or group 2.
Subjects in group 1 receive treatment A in the first period and then receive treatment B in the 
second period. Subjects in group 2 receive treatment B in the first period and then receive treatment A in the 
second period. This design is efficient under the assumption 
that there is no differential carryover effect. It is not an appropriate design unless there is strong
prior information that this assumption holds. 
However, a commonly occurring scenario is that it
is not certain that this assumption holds. We consider this scenario.
To deal with this uncertainty,
it has been suggested (starting with Grizzle, 1965, 1974 and endorsed by Hills and Armitage, 1979) that a 
preliminary test of the null hypothesis that this assumption holds be carried 
out before proceeding with further inference. 
If this test leads to acceptance of this null hypothesis then further
inference proceeds on the basis that it was known
{\it a priori} that there is no differential carryover effect. If, on the other hand, this null hypothesis
is rejected then further inference is based solely on data from the first period,
since this is unaffected by any carryover effect.
In a landmark paper, Freeman (1989) 
showed that the use of 
such a preliminary hypothesis test prior to the construction of a confidence interval 
with nominal coverage $1-\alpha$ 
leads to a confidence interval with minimum coverage probability far below $1-\alpha$.
For simplicity, Freeman supposes that the subject variance and the error variance are known.
In other words, Freeman presents a large-sample analysis.
Freeman's conclusion that the use of a preliminary test in this way
`is too potentially misleading to be of practical use' is now widely accepted (Senn, 2006).
Freeman's finding is consistent with the known deleterious effect of preliminary 
hypothesis tests on the coverage properties of 
subsequently-constructed confidence intervals in the context of a linear regression
with independent and identically distributed zero-mean normal errors (see e.g. 
Kabaila, 2005; Kabaila and Leeb, 2006; Giri and Kabaila, 2008; Kabaila and Giri, 2008).

A Bayesian analysis that incorporates prior information about the differential carryover effect
is provided by Grieve (1985, 1986). However, there is currently no valid frequentist confidence 
interval for the difference $\theta$ of the two treatment effects that utilizes the 
uncertain prior information that there is no differential carryover effect. 
Similarly to Hodges and Lehmann (1952), Bickel (1983, 1984), Kabaila (1998), Kabaila and Giri (2007ab),
Farchione and Kabaila (2008) and Kabaila and Tuck (2008), our aim is to utilize the uncertain prior
information in the frequentist inference of interest,
whilst providing a safeguard in case this prior information happens to be incorrect.
We follow Freeman (1989) and assume that the between-subject variance and the error variance are known.
As already noted, this corresponds to a large-sample analysis.
The usual $1-\alpha$ confidence interval for $\theta$ based solely
on data from the first period is unaffected by any differential carryover effect.
We use this interval as the standard against which other 
$1-\alpha$ confidence intervals will be assessed. We therefore call this confidence
interval the `standard $1-\alpha$ confidence interval'. We assess a 
$1-\alpha$ confidence interval for $\theta$ using the ratio
\begin{equation*}
\frac{\text{expected length of this confidence interval}}
{\text{length of the
standard $1-\alpha$ confidence interval}.}
\end{equation*}
We call this ratio the scaled expected length of this confidence interval.
We find a new $1-\alpha$ confidence interval that utilizes the uncertain prior information 
that the differential carryover effect is zero, in the following sense.
This new interval has scaled expected length that 
(a) is substantially less than 1 when the prior information that there is no differential
carryover effect holds
and (b) has a maximum value that is not too large. Also, this confidence interval
coincides with the standard $1-\alpha$ confidence interval
when the data strongly contradicts the prior information that there is no differential carryover
effect. Additionally, this confidence
interval has the attractive feature that it has endpoints that are
continuous functions of the data.

The properties of the new large-sample confidence interval, described in Section 2,
are illustrated in Section 3 by a detailed analysis
of the case that the between-subject variance and the error variance are equal
and $1-\alpha = 0.95$.
In Section 2, we define the parameter $\gamma$ to be the differential carryover divided by
the standard deviation of the least squares estimator of the differential carryover. 
As proved in Section 2, the coverage probability of the new confidence interval for $\theta$ is an even function
of $\gamma$. The top panel of Figure 2 is a plot of  the coverage probability of the new 0.95 confidence interval for $\theta$
as a function of $\gamma$. This plot shows that the new 0.95 confidence interval for $\theta$ has
coverage probability 0.95 throughout the
parameter space. As proved in Section 2, the scaled expected length of the new confidence interval for $\theta$ is an even function
of $\gamma$. The bottom panel of Figure 2 is a plot of the square of the scaled expected length of the 
new 0.95 confidence interval for $\theta$
as a function of $\gamma$. When the prior information is correct (i.e. $\gamma=0$), 
we gain since the square of the scaled expected length
is substantially smaller than 1. The maximum value of the square of the scaled expected length is not too large.
The new 0.95 confidence interval for $\theta$ coincides with the standard $1-\alpha$ confidence interval
when the data strongly contradicts the prior information. This is reflected in Figure 2 by the fact that the
square of the scaled expected length approaches 1 as $\gamma \rightarrow \infty$.

In Section 4, we compare the two-period crossover trial with a completely randomized design
with the same number of measurements of response, using a
large sample analysis. 
We assume that the new 0.95 confidence interval is used to summarise the data from the two-period
crossover trial. We show that 
the uncertainty in the prior information that there is no differential carryover
effect has the following consequence. Subject to a reasonable upper bound on how badly the 
new 0.95 confidence interval 
can perform relative to the usual 0.95 confidence interval for $\theta$ based on data from the 
completely randomized design, the completely randomized design is better than the two-period crossover trial
for all (subject variance)/(error variance) $\in (0, 11.6263]$.
In Section 5 we describe the implications for finite samples of the results described in
Sections 3 and 4.

\bigskip

\noindent {\bf 2. New large-sample confidence interval utilizing prior information about the differential
carryover effect}

\medskip

We assume the model for the two-treatment two-period crossover trial put forward by 
Grizzle (1965), as described by Grieve (1987). Let $n_1$ and $n_2$ denote the number 
of subjects in groups 1 and 2 respectively. Also let $Y_{ijk}$ denote the response of
the $j$th subject in the $i$th group and the $k$th period ($i=1,2$; $j=1,\ldots,n_i$;
$k=1,2$). The model is
\begin{equation*}
Y_{ijk} = \mu + \xi_{ij} + \pi_k + \phi_{\ell} + \lambda_{\ell} + \varepsilon_{ijk}
\end{equation*}
where $\mu$ is the overall population mean, $\xi_{ij}$ is the effect of the $j$th 
patient in the $i$th group, $\pi_k$ is the effect of the $k$th period, $\phi_{\ell}$
is the effect of the $\ell$th treatment,
$\lambda_{\ell}$ is the residual effect of the $\ell$th treatment
 and $\varepsilon_{ijk}$ is the random error.
We assume that the $\xi_{ij}$ and $\varepsilon_{ijk}$ are independent and that the 
$\xi_{ij}$ are identically $N(0,\sigma_s^2)$ distributed and the $\varepsilon_{ijk}$ 
are identically $N(0,\sigma_{\varepsilon}^2)$ distributed,
where $\sigma_s^2 > 0$ and $\sigma_{\varepsilon}^2 > 0$.
Let $m = (1/n_1)+(1/n_2)$, $\sigma^2 = \sigma_{\varepsilon}^2 + \sigma_s^2$ and
$\rho = \sigma_s^2 / (\sigma_{\varepsilon}^2 + \sigma_s^2)$. The parameter of 
interest is $\theta = \phi_1 - \phi_2$. The parameter describing the differential 
carryover effect is $\psi = (\lambda_1 - \lambda_2)/2$. 
We suppose that we have uncertain prior information that $\psi = 0$.

We use the notation $\bar Y_{i \boldsymbol{\cdot} \, k} = (1/n_i) \sum_{j=1}^{n_1} Y_{ijk}$
($i=1,2$).
Our statistical analysis will be described entirely in terms of the following random variables:
$A = \big(\bar Y_{1 \boldsymbol{\cdot} 1} - \bar Y_{1 \boldsymbol{\cdot} 2} - \bar Y_{2 \boldsymbol{\cdot} 1}
+ \bar Y_{2 \boldsymbol{\cdot} 2} \big)/2$,
$\hat \Psi = \big(\bar Y_{1 \boldsymbol{\cdot} 1} + \bar Y_{1 \boldsymbol{\cdot} 2} - \bar Y_{2 \boldsymbol{\cdot} 1}
- \bar Y_{2 \boldsymbol{\cdot} 2} \big)/2$,
\begin{align*}
%A &= \big(\bar Y_{1 \boldsymbol{\cdot} 1} - \bar Y_{1 \boldsymbol{\cdot} 2} - \bar Y_{2 \boldsymbol{\cdot} 1}
%+ \bar Y_{2 \boldsymbol{\cdot} 2} \big)/2 \\ 
%\hat \Psi &= \big(\bar Y_{1 \boldsymbol{\cdot} 1} + \bar Y_{1 \boldsymbol{\cdot} 2} - \bar Y_{2 \boldsymbol{\cdot} 1}
%- \bar Y_{2 \boldsymbol{\cdot} 2} \big)/2 \\ 
V &= \frac{1}{2} \left ( \sum_{j=1}^{n_1} \left ( ( Y_{1j1} - \bar Y_{1 \boldsymbol{\cdot} 1}) -
( Y_{1j2} - \bar Y_{1 \boldsymbol{\cdot} 2}) \right)^2 + 
\sum_{j=1}^{n_2} \left ( ( Y_{2j1} - \bar Y_{2 \boldsymbol{\cdot} 1}) -
( Y_{2j2} - \bar Y_{2 \boldsymbol{\cdot} 2}) \right)^2 \right),\\
W &= \frac{1}{2} \left ( \sum_{j=1}^{n_1} \left ( ( Y_{1j1} - \bar Y_{1 \boldsymbol{\cdot} 1}) +
( Y_{1j2} - \bar Y_{1 \boldsymbol{\cdot} 2}) \right)^2 + 
\sum_{j=1}^{n_2} \left ( ( Y_{2j1} - \bar Y_{2 \boldsymbol{\cdot} 1}) +
( Y_{2j2} - \bar Y_{2 \boldsymbol{\cdot} 2}) \right)^2 \right).
\end{align*}
These random variables are independent and they have the following distributions:
$A \sim N \big(\theta - \psi, m \sigma_{\varepsilon}^2 / 2 \big)$, 
$\hat \Psi \sim N \big(\psi, m (\sigma_{\varepsilon}^2 + 2 \sigma_s^2)/2 \big)$,
$V/\sigma_{\varepsilon}^2 \sim \chi_{n_1 + n_2-2}^2$ and
$W/(\sigma_{\varepsilon}^2 + 2 \sigma_s^2) \sim \chi_{n_1 + n_2-2}^2$. 
%
%\begin{align*}
%&A \sim N \big(\theta - \psi, m \sigma_{\varepsilon}^2 / 2 \big) \\
%&\hat \Psi \sim N \big(\psi, m (\sigma_{\varepsilon}^2 + 2 \sigma_s^2)/2 \big) \\
%&\frac{V}{\sigma_{\varepsilon}^2} \sim \chi_{n_1 + n_2-2}^2 \\
%&\frac{W}{\sigma_{\varepsilon}^2 + 2 \sigma_s^2} \sim \chi_{n_1 + n_2-2}^2 
%\end{align*}
%
Define $\hat \Theta = A + \hat \Psi = \bar Y_{1 \boldsymbol{\cdot} 1} - \bar Y_{2 \boldsymbol{\cdot} 1}$.
This estimator of $\theta$ is based solely on the data from period 1. Consequently, it is 
not influenced by any carryover effects.
Note that
\begin{equation}
\label{original_model}
\left[\begin{matrix} \hat \Theta\\ \hat \Psi \end{matrix}
\right] \sim N \left ( \left[\begin{matrix} \theta \\ \psi \end{matrix}
\right], \sigma^2  \left[\begin{matrix} m \quad \quad m \tilde \rho^2 \\ 
m \tilde \rho^2 \quad m \tilde \rho^2 \end{matrix}
\right] \right ).
\end{equation}
where $\tilde \rho$ denotes the correlation between $\hat \Theta$ and $\hat \Psi$
and is equal to $\sqrt{(1+\rho)/2}$.

We follow Freeman (1989) and assume that the subject variance $\sigma_s^2$
and the error variance $\sigma_{\varepsilon}^2$ are known. This implies that the
parameters $\sigma^2$ and $\tilde \rho$ are known. Using the random variables $V$ and
$W$ in the obvious way, $\sigma_s^2$ and $\sigma_{\varepsilon}^2$ can be estimated
consistently as $n_1 + n_2 \rightarrow \infty$. In other words, we are using a 
large-sample analysis.

We use the notation $[a \pm b]$ for the interval $[a-b, a+b]$ ($b > 0$). Define
$c_{\alpha} = \Phi^{-1}(1-\frac{\alpha}{2})$, where $\Phi$ denotes the $N(0,1)$ 
cumulative distribution function. The usual $1-\alpha$ confidence interval for
$\theta$, based solely on data from the first period, is
$I = \big[\hat \Theta \pm c_{\alpha} \sqrt{m} \sigma \big]$.
%
%\begin{equation*}
%I = \big[\hat \Theta \pm c_{\alpha} \sqrt{m} \sigma \big].
%\end{equation*}
%
Define the following confidence interval
for $\theta$:
\begin{equation*}
\label{J(b,s)}
J(b,  s) = \bigg [ \hat \Theta -
\sqrt{m} \sigma \, b\bigg(\frac{\hat{\Psi}}{\sigma \sqrt{m} \tilde \rho } \bigg) \, \pm \,
\sqrt{m} \sigma \, s\bigg(\frac{|\hat{\Psi}|}{\sigma \sqrt{m} \tilde \rho } \bigg)
\bigg ],
\end{equation*}
where the functions $b$ and $s$ are required to satisfy the following restriction.
\smallskip

\baselineskip=20pt

\noindent {\underbar{Restriction 1}} \ 
$b: \mathbb{R} \rightarrow \mathbb{R}$  is an odd function and $s: [0, \infty)
\rightarrow [0, \infty)$.

\smallskip

\noindent Invariance arguments, of the type used by Farchione and Kabaila (2008),
may be used to motivate this restriction. For the sake of brevity, these arguments
are omitted. We also require the functions $b$ and $s$ to satisfy the following 
restriction.

\smallskip

\noindent {\underbar{Restriction 2}} \ 
$b$ and $s$ are continuous functions.

\smallskip
\noindent This implies that the endpoints of the confidence interval $J(b,s)$ are continuous functions
of the data. Finally, we require the confidence interval $J(b,s)$ to coincide with
the standard $1-\alpha$ confidence interval $I$ when the data strongly contradict
the prior information. The statistic $|\hat \Psi|/(\sigma \sqrt{m} \tilde \rho)$ provides some indication
of how far away $\psi / (\sigma \sqrt{m} \tilde \rho)$ is from 0.
We therefore require that 
the functions $b$ and $s$ 
satisfy the following restriction.
\smallskip

\noindent {\underbar{Restriction 3}} \
$b(x)=0$ for all $|x| \ge d$ and $s(x)=c_\alpha$ for all $x > d$,
where $d$ is a (sufficiently large) specified positive number.

\smallskip

Define $\gamma = \psi/(\sigma \sqrt{m} \tilde \rho)$,
$G = (\hat \Theta - \theta)/(\sigma \sqrt{m})$
and $H = \hat \Psi/(\sigma \sqrt{m} \tilde \rho)$. 
It follows from 
\eqref{original_model} that
\begin{equation}
\label{model_G_H}
\left[\begin{matrix} G\\ H \end{matrix}
\right] \sim N \left ( \left[\begin{matrix} 0 \\ \gamma \end{matrix}
\right], \left[\begin{matrix} 1 \quad \tilde \rho\\ \tilde \rho \quad 1 \end{matrix}
\right] \right ).
\end{equation}
It is straightforward to show that
the coverage probability $P\big( \theta \in J(b, s) \big)$
is equal to
$P \big ( \ell(H) \le G \le u(H) \big )$
where $\ell(h) = b(h) - s(|h|)$ and $u(h) = b(h) + s(|h|)$.
For given $b$, $s$ and $\tilde \rho$, this coverage probability is a function of $\gamma$.
We denote this coverage probability by $c(\gamma;b,s, \tilde \rho)$.

Part of our evaluation of the confidence interval $J(b,s)$ consists of comparing it with the
standard $1-\alpha$ confidence interval
$I$ using the criterion
\begin{equation}
\label{criterion_initial}
\frac{\text{expected length of $
J(b, s)$}}
{\text{length of $I$}}.
\end{equation}
We call this the scaled expected length of $J(b, s)$. This is equal to
$E(s(|H|))/c_{\alpha}$. 
This is a function of $\gamma$ for given $s$. We denote this function by $e(\gamma;s)$.
Clearly, for given $s$, $e(\gamma;s)$ is an even function of $\gamma$.

Our aim is to find functions $b$ and $s$ that satisfy Restrictions 1--3 and 
such that (a) the minimum of $c(\gamma;b,s, \tilde \rho)$ over $\gamma$ is
$1-\alpha$ and (b)
\begin{equation}
\label{criterion}
\int_{-\infty}^{\infty} (e(\gamma;s) - 1) \, d \nu(\gamma)
\end{equation}
is minimized, where the weight function $\nu$ has been chosen to be
\begin{equation*}
\label{mixed_wt_fn}
\nu(x) = \omega x + {\cal H}(x) \ \text{ for all } \ x \in \mathbb{R},
\end{equation*}
where $\omega$ is a specified nonnegative number and
${\cal H}$ is the unit step function defined by ${\cal H}(x) = 1$ for $x \ge 0$ 
and ${\cal H}(x) = 0$ for $x < 0$. 
The larger the value of $\omega$, the smaller the relative weight given to
minimizing $e(\gamma;s)$ for $\gamma=0$, as opposed to
minimizing $e(\gamma;s)$ for other values of $\gamma$.

The following theorem (cf. Kabaila and Giri, 2007a) provides computationally convenient expressions
for the coverage probability and scaled expected length of $J(b, s)$.

\medskip

\noindent {\bf Theorem 2.1}

\noindent (a) {\sl Define the functions
$k^{\dag}(h, \gamma, \tilde \rho) = \Lambda \left( -c_{\alpha}, c_{\alpha};
 \tilde \rho(h-\gamma), 1-\tilde \rho^2 \right )$ and
$k(h,\gamma, \tilde \rho) = \Lambda \left(\ell(h), u(h); \tilde \rho(h-\gamma),1- \tilde \rho^2 \right )$,
where $\Lambda(x, y; \mu, v) = P(x \le Z \le y)$ for $Z \sim N(\mu,v)$.
The coverage probability $P\big( \theta \in J(b, s) \big)$ is equal to
\begin{equation}
%\tag{C.4}
\label{cov_prob}
(1-\alpha) +
\int_{-d}^{d} \big( k(h, \gamma, \tilde \rho) - k^{\dag}(h, \gamma, \tilde \rho) \big)
\, \phi(h-\gamma)\,  dh,
\end{equation}
where $\phi$ denotes the $N(0,1)$ probability density function.
For given $b$, $s$ and $\tilde \rho$, $c(\gamma;b,s, \tilde \rho)$ is an even function of $\gamma$.}

\smallskip

\noindent (b) {\sl The scaled expected length of $J(b, s)$
is}
\begin{equation}
%\tag{B.3}
\label{comp_conv_e}
e(\gamma;s) = 1 +
 \frac{1} {c_{\alpha}}
 \int^{d}_{- d} \left (s(|h|) - c_{\alpha} \right )
\phi(h -\gamma) \, dh.
\end{equation}

\medskip

Substituting \eqref{comp_conv_e} into \eqref{criterion} we obtain that \eqref{criterion} is equal to
\begin{equation}
\label{criterion_final}
 \frac{1} {c_{\alpha}}  \int_{-\infty}^{\infty}
 \int^{d}_{- d} \left (s(|h|) - c_{\alpha} \right )
\phi(h -\gamma) \, dh \, d \nu (\gamma) 
=\frac{2} {c_{\alpha}}
 \int^{d}_{0} \left (s(h) - c_{\alpha} \right )
(\omega + \phi(h)) \, dh. 
\end{equation}

For computational feasibility, we specify the following parametric forms for the
functions $b$ and $s$.
We require $b$ to be a continuous function and so it is necessary that
$b(0)=0$.
Suppose that $x_1, \ldots, x_q$
satisfy $0 = x_1 < x_2 < \cdots < x_q = d$.
Obviously, $b(x_1)=0$, $b(x_q)=0$ and $s(x_q)=c_{\alpha}$.
The function $b$ is fully specified by the vector $\big (b(x_2), \ldots, b(x_{q-1}) \big)$
as follows. Because $b$ is assumed to be an odd function, we know that
$b(-x_i) = -b(x_i)$ for $i=2,\ldots,q$.
We specify the value of $b(x)$ for any $x \in [-d,d]$
by cubic spline interpolation for these given function values, subject to the constraint
that $b^{\prime}(-d)=0$ and $b^{\prime}(d)=0$.
We fully specify the function $s$ by the vector $\big (s(x_1), \ldots, s(x_{q-1}) \big)$
as follows.
The value of $s(x)$ for any $x \in [0,d]$ is specified
by cubic spline interpolation for these given function values (without any endpoint conditions
on the first derivative of $s$). We call $x_1, x_2, \ldots x_q$ the knots.

To conclude this section, the new $1-\alpha$ confidence interval for $\theta$ that utilizes the prior information
that $\psi=0$ is obtained as follows. For a judiciously-chosen set of values
of $d$, $\omega$ and knots $x_i$, we carry out the following computational procedure.

\smallskip

\noindent {\underbar{Computational Procedure}} \
Compute the functions $b$ and $s$, satisfying Restrictions 1--3 and taking the 
parametric forms described above, such that (a) the minimum over $\gamma \ge 0$ of
\eqref{cov_prob} is $1-\alpha$ and (b) the criterion \eqref{criterion_final} is minimized.
Plot $e^2(\gamma;s)$, the square of the scaled expected length,
as a function of $\gamma \ge 0$.

\smallskip

\noindent Based on these plots and the strength of our prior information that $\psi=0$,
we choose appropriate values of $d$, $\omega$ and knots $x_i$.
The confidence interval corresponding to this choice is the new
$1-\alpha$ confidence interval for $\theta$.
For given $\omega$, the functions $b$ and $s$ can be chosen to be functions of 
$\tilde \rho$, since $\tilde \rho$ is assumed to be known.
All the computations for the present paper were performed with programs written in MATLAB,
using the Optimization and Statistics toolboxes.

\bigskip

\noindent {\bf 3. Illustration of the properties of the new confidence interval}

\medskip

The parameter $\tilde \rho$ lies in the interval $\big(1/\sqrt{2},\, 1 \big)$. To illustrate the properties
of the new $1-\alpha$ confidence interval for $\theta$, consider the case that 
$\sigma_s^2/\sigma_{\varepsilon}^2=1$, so that $\tilde \rho = \sqrt{3}/2$. Suppose that
$1-\alpha = 0.95$. We have followed the Computational Procedure, described in the previous
section, with $d=6$, $\omega=0.2$ and evenly-spaced knots at $0, 6/8, \ldots,6$. The 
resulting functions $b$ and $s$, which specify the new 0.95 confidence interval 
for $\theta$, are plotted in Figure 1. The performance of this confidence intervals
is shown in Figure 2. When the prior information is correct (i.e. $\gamma=0$),
we gain since $e^2(0;s)=0.8527$.
The maximum value of $e^2(\gamma;s)$ is
1.1239. This confidence interval coincides with the standard
$1-\alpha$ confidence interval for $\theta$ when the data strongly
contradicts the prior information, so that $e^2(\gamma;s)$
approaches 1 as $\gamma \rightarrow \infty$.

The value of $\omega=0.2$ was obtained
from the following search. Consider 
$\omega =
0.05$, 0.2 , 0.5 and 1. The Computational Procedure was applied
for each of these values. As expected from the form of the weight
function, for each of these values of $\omega$, $e^2(\gamma;s)$
is minimized at $\gamma=0$. For a given value of $\lambda$, define
the `expected gain' to be $\big(1 - e^2(0;s) \big)$ and the
`maximum potential loss' to be $\big (\max_{\gamma} e^2(\gamma;s)
- 1 \big)$. As shown in Table 1, as $\omega$ increases (a) the
expected gain decreases and (b) the ratio (expected gain)/(maximum
potential loss) increases. By choosing $\omega=0.2$ we have both
a reasonably large expected gain and a reasonably large value of
the ratio (expected gain)/(maximum potential loss).

\medskip

\begin{table}[h]
\renewcommand{\arraystretch}{1.3}
\hfil
\begin{tabular}{|c|c|c|c|c|c|c|c|c|}
  \hline
  % after \\: \hline or \cline{col1-col2} \cline{col3-col4} ...
  $\omega$ & 0.05 & 0.2 & 0.5& 1 \\
  \hline
  expected gain & 0.2173 & 0.1473 & 0.0904& 0.0542   \\
  \hline
 maximum potential loss & 0.2982 & 0.1239 & 0.0595 & 0.0324 \\
  \hline
  (expected gain)/(maximum potential loss) & 0.7288 & 1.1892 & 1.5206 & 1.6704 \\
  \hline
\end{tabular}
\hfil \caption{Performance of the 0.95 confidence interval for
$d=6$ and evenly-spaced knots at $0, 6/8, \ldots,6$,
when we vary over
$\omega \in \{0.05, 0.2, 0.5, 1\}$.} \label{size_table}
\end{table}

\baselineskip=20pt

\noindent {\bf 4. Comparison of the two-period crossover trial with a completely \newline randomized design
with the same number of measurements of response}

\medskip

For the two-period crossover trial, the total number of measurements of response is $2M$, where
$M = n_1 + n_2$. Following Brown (1980), we compare this design with a completely randomized design with the same total number
of measurements of the response. For the completely randomized design, we have $M$ randomly-chosen subjects
given treatment A and $M$ randomly-chosen subjects
given treatment B. Let $Y_1^A, \ldots, Y_M^A$ denote the responses for the $M$ subjects given
treatment A and let $Y_1^B, \ldots, Y_M^B$ denote the responses for the $M$ subjects given
treatment B. A model for these responses that is consistent with the model used for the two-period
crossover trial is the following. Suppose that $Y_1^A, \ldots, Y_M^A, Y_1^B, \ldots, Y_M^B$ are
independent random variables, with $Y_1^A, \ldots, Y_M^A$ identically $N(\phi_1, \sigma^2)$ distributed
and $Y_1^B, \ldots, Y_M^B$ identically $N(\phi_2, \sigma^2)$ distributed. The usual estimator of 
$\theta = \phi_1 - \phi_2$ is
$\widetilde \Theta = \big((Y_1^A + \cdots + Y_M^A) - (Y_1^B + \cdots + Y_M^B)\big)/M$.
Obviously, $\widetilde \Theta \sim N(\theta, 2\sigma^2/M)$.

\baselineskip=21pt

\FloatBarrier

%\vspace{25cm}
%\medskip

%\FloatBarrier
\begin{figure}[h]
\label{Figure1} \centering
\includegraphics[scale=1]{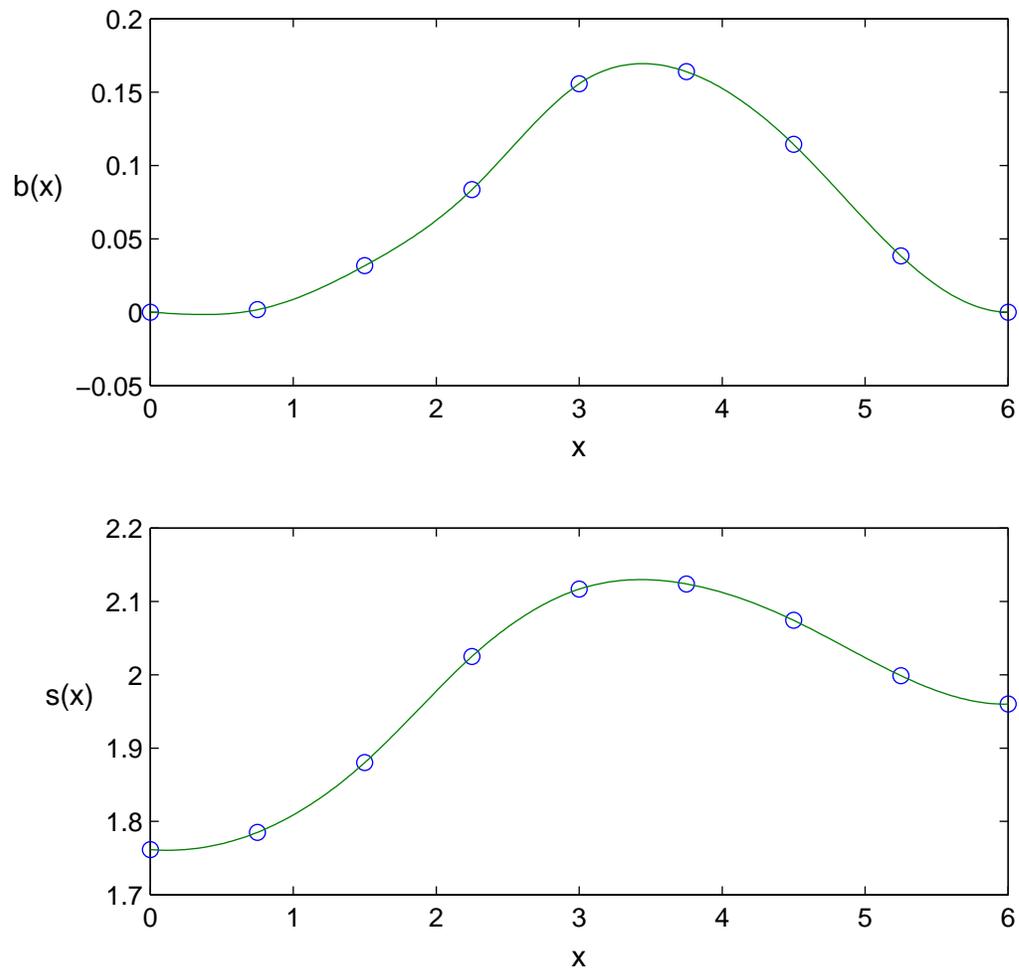}
       \caption{ Plots of the functions $b$ and $s$ for $\sigma_s^2/\sigma_{\varepsilon}^2=1$ and $1-\alpha=0.95$.
       These functions were obtained using $d=6$, $\omega=0.2$ and evenly-spaced knots $x_i$ at $0, 6/8, \ldots,6$.}
\end{figure}
%\FloatBarrier

%\FloatBarrier
\begin{figure}[h]
\label{Figure1} \centering
\includegraphics[scale=1]{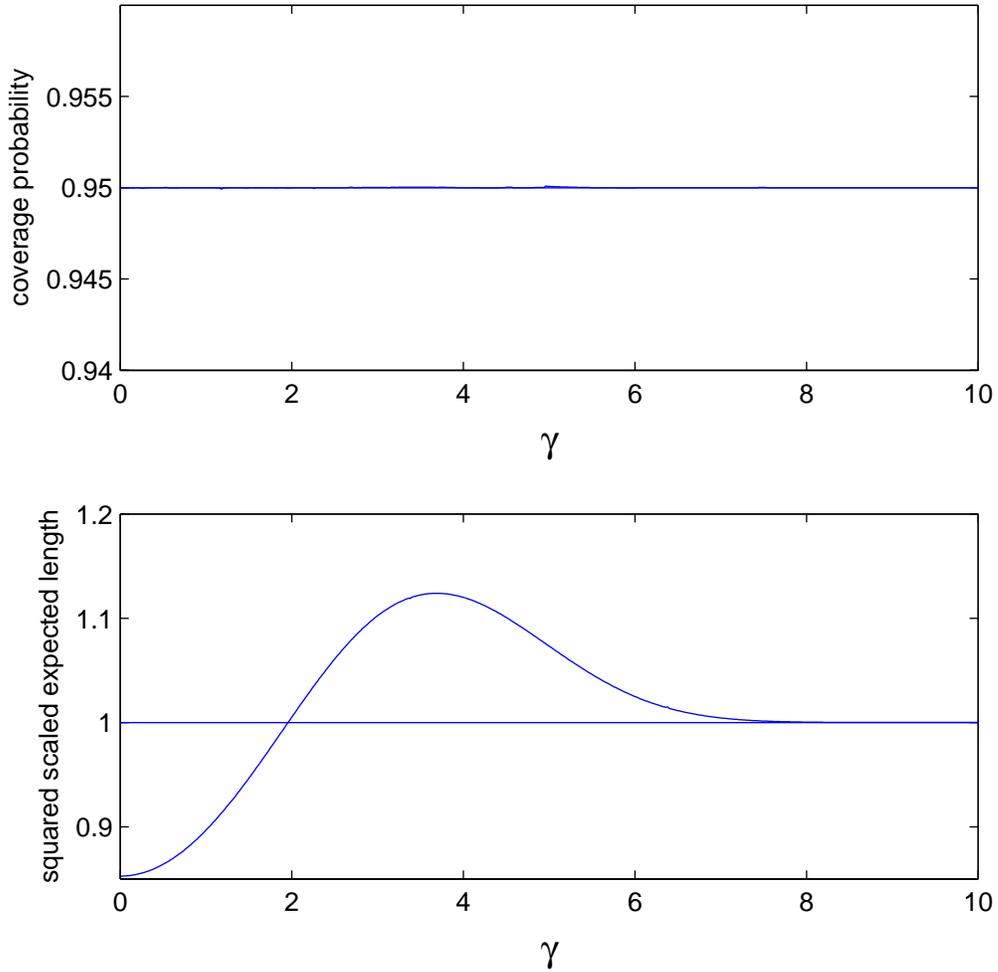}
       \caption{ Plots of the coverage probability and $e^2(\gamma;s)$, the squared scaled expected length,
{\Big(}as functions of $\gamma = \psi/\sqrt{\text{var}(\hat \Psi)}${\Big )} of the new 0.95
confidence interval for $\theta$ 
when $\sigma_s^2/\sigma_{\varepsilon}^2=1$.
These functions were obtained using $d=6$, $\omega=0.2$ and evenly-spaced knots $x_i$ at $0, 6/8, \ldots,6$.}
\end{figure}
\FloatBarrier

\bigskip

Now, following Brown (1980), consider the case that there is no differential carryover effect i.e.
that $\psi=0$. In this case, $\theta$ is estimated by $A \sim N(\theta, m \sigma_{\varepsilon}^2/2)$.
Thus 
\begin{equation*}
\frac{\text{var}(A)}{\text{var}(\widetilde \Theta)} = \frac{n_1 + n_2}{4} 
\left(\frac{1}{n_1} + \frac{1}{n_2} \right) \frac{1}{1 + (\sigma_s^2/\sigma_{\varepsilon}^2)}.
\end{equation*}
As this expression shows, the efficiency of the two-period crossover trial, relative to the
completely randomized design, is an increasing function of $\sigma_s^2/\sigma_{\varepsilon}^2$.
For the case $n_1 = n_2 = n$, the two-period crossover trial is more efficient than the
completely randomized design for all $\sigma_s^2/\sigma_{\varepsilon}^2 > 0$.
In other words, if we are absolutely certain that there is no differential carryover effect
then we should always use the two-period crossover trial, as opposed to the 
completely randomized design.

However, as noted in the introduction, it is commonly the case that it is not certain 
that there is no differential carryover effect. We ask the following question.
What is the efficiency of the two-period crossover trial relative to the  
completely randomized design in this case? We consider this question in the context that 
$\sigma_s^2$ and $\sigma_{\varepsilon}^2$ are known. In other words, we consider
this question in the context of large samples. 
We also assume that the new $1-\alpha$ confidence interval described in Section 2
is used to summarise the data from
the two-period crossover trial. For simplicity suppose
that $n_1 = n_2 = n$. Based on data from a completely randomized design, that usual
$1-\alpha$ confidence interval for $\theta$ is 
$K = \big[ \widetilde \Theta \pm c_{\alpha} \sigma/\sqrt{n} \big]$.
In earlier sections, we have assessed the new $1-\alpha$ confidence interval 
using the scaled expected length criterion \eqref{criterion_initial}, denoted by $e(\gamma;s)$.
To compare the two-period crossover trial with a completely randomized design with the same total
number of measurements, we now use the criterion
\begin{equation*}
r(\gamma;s) = 
\frac{\text{expected length of $
J(b, s)$}}
{\text{length of $K$}}.
\end{equation*}
Note that $r(\gamma;s) = \sqrt{2} \, e(\gamma;s)$, so that 
$r^2(\gamma;s) = 2 \, e^2(\gamma;s)$. For a given value of 
$\tilde \rho \in (1/\sqrt{2}, 1)$, let us restrict attention to the class
${\cal C}(\tilde \rho, 1-\alpha)$ of new $1-\alpha$ confidence intervals 
that satisfy the constraint
$\max_{\gamma} e^2(\gamma;s) \le 1.25$,
so that 
\begin{equation}
\label{upper_bnd}
\max_{\gamma} r^2(\gamma;s) \le 1.5.
\end{equation}
This condition puts an upper bound on how badly the new $1-\alpha$ confidence interval 
can perform relative to the confidence interval $K$ based on data from the 
completely randomized design.
Consider the particular case that $1-\alpha = 0.95$. 
For each $\tilde \rho \in (1/\sqrt{2}, 0.98]$, i.e. for each 
$\sigma_s^2/\sigma_{\varepsilon}^2 \in (0, 11.6263]$,
we find computationally that $\min_{\gamma} r^2(\gamma;s) > 1$ for every
new $1-\alpha$ confidence interval belonging to ${\cal C}(\tilde \rho, 0.95)$.
In other words, if we impose the reasonable constraint \eqref{upper_bnd}
then, for $1-\alpha = 0.95$ and large samples, the completely randomized design
is better than the two-period crossover trial for each 
$\sigma_s^2/\sigma_{\varepsilon}^2 \in (0, 11.6263]$. This is a complete contrast
to the case that we are absolutely certain that there is no differential carryover effect.

\bigskip

\noindent {\bf 5. Implications for finite samples}

\medskip

By replacing the parameters $\sigma$ and $\tilde \rho$ by their obvious estimators
(based on the statistics $V$ and $W$) in the new large-sample $1-\alpha$ confidence interval
described in Section 2, we obtain a new finite-sample confidence interval for $\theta$.
This new finite-sample confidence interval will have coverage and scaled expected length
properties that will approach the corresponding properties for the new large-sample $1-\alpha$ confidence interval
as $n_1 + n_2 \rightarrow \infty$. This suggests that it will be possible to design confidence
intervals for $\theta$ that utilize the uncertain prior information that there is no differential
carryover effect for small and medium, as well as large sample sizes. This also suggests that the result found
in Section 4 will also be reflected in small and medium, as well as large samples sizes. 
We expect that subject 
to a reasonable upper bound on how badly any  
new finite-sample 0.95 confidence interval 
can perform relative to the usual 0.95 confidence interval for $\theta$ based on data from the 
completely randomized design, the completely randomized design is better than the two-period crossover trial
for a very wide range of values of (subject variance)/(error variance).

\bigskip

\noindent {\bf {Appendix. Proof of Theorem 2.1}}

\medskip

In this appendix we prove Theorem 2.1.

\smallskip

\noindent {\bf Proof of part (a).} \
It follows from \eqref{model_G_H}
that the probability density function of $H$, evaluated at $h$, is $\phi(h-\gamma)$. Thus
\begin{equation}
\tag{A.1}
\label{A.1}
 c(\gamma;b,s,\tilde \rho)
 = \int_{-\infty}^{\infty} \int_{\ell(h)}^{u(h)}
f_{G|H}(g|h) \, dg \, \phi(h-\gamma)\,  dh 
\end{equation}
where $f_{G|H}(g|h)$ denotes the probability density function of $G$ conditional on $H=h$, evaluated at $g$.
The probability distribution of $G$ conditional on $H=h$ is
$N \big( \tilde \rho (h - \gamma), 1 - \tilde{\rho}^2 \big )$. Thus the right hand side of
\eqref{A.1} is equal to
\begin{equation}
\tag{A.2}
\label{A.2}
\int_{-\infty}^{\infty} k(h, \gamma, \tilde \rho)
\, \phi(h-\gamma)\,  dh 
\end{equation}
The standard $1-\alpha$ confidence interval $I$ has coverage probability $1-\alpha$.
Hence
\begin{equation}
\tag{A.3}
\label{A.3}
1-\alpha = \int_{-\infty}^{\infty} k^{\dag}(h, \gamma, \tilde \rho)
\, \phi(h-\gamma)\,  dh.
\end{equation}
The result follows from subtracting \eqref{A.3} from \eqref{A.2} and noting that
$b(x)=0$ for all $|x| \ge d$ and $s(x)=c_{\alpha}$ for all $x \ge d$.
By a consideration of the distribution of $(-G,-H)$, it may be shown that
$c(\gamma;b,s, \tilde \rho)$ is
an even function of $\gamma$, for given $b$, $s$ and $\tilde \rho$.

\smallskip

\noindent{\bf Proof of part (b).} The result is an immediate consequence of the
fact that $b(x)=0$ for all $|x| \ge d$ and $s(x)=c_{\alpha}$ for all $x \ge d$.

\bigskip

\baselineskip=19pt

\noindent {\bf References}

\medskip

\rf Bickel, P.J., 1983. Minimax estimation of the mean of a normal
distribution subject to doing well at a point. In:  Rizvi, M.H.,
Rustagi, J.S., Siegmund, D., (Eds), Recent Advances in Statistics,
Academic Press, New York, 511--528.

\rf Bickel, P.J., 1984. Parametric robustness: small biases can be
worthwhile. Annals of Statistics 12, 864--879.

\rf Brown, B.W., 1980. The crossover experiment in clinical trials.
Biometrics 36, 69--79.

%\rf Bickel, P.J., Doksum, K.A. (1977). Mathematical Statistics,
%Basic Ideas and Selected Topics. Holden-Day, Oakland, California.

%\rf Bohrer, R., Sheft, J., 1979. Misclassification probabilities
%in $2^3$ factorial experiments. Journal of Statistical Planning
%and Inference 3, 79--84.

%\rf Buring, J.E., Hennekens, C.H., 1990. Cost and efficiency in
%clinical trials: the U.S. Physicians' Health Study. Statistics in
%Medicine 9, 29--33.

%\rf Casella, G., Berger, R. L., 2002. Statistical Inference, 2nd
%ed. Duxbury, Pacific Grove, California.

%\rf Dub\'e, M.A., Penlidis, A., Reilly, P.M., 1996. A systematic
%approach to the study of multicomponent polymerization kinetics:
%the butyl acrylate/methyl methacrylate/vinyl acetate example. IV.
%Optimal Bayesian design of emulsion terpolymerization experiments
%in a pilot plant reactor. Journal of Polymer Science: Part A:
%Polymer Chemistry 34, 811--831.

%\rf Fabian, V., 1991. On the problem of interactions in the
%analysis of variance. Journal of the American Statistical
%Association 86, 362--373.

\rf Farchione, D., Kabaila, P., 2008. Confidence intervals for the
normal mean utilizing prior information. Statistics
\& Probability Letters 78, 1094--1100.

\rf Freeman, P.R., 1989. The performance of the two-stage analysis
of two-treatment, two-period crossover trials. Statistics in Medicine
8, 1421--1432.

\rf Giri, K., Kabaila, P., 2008. The coverage probability of
confidence intervals in $2^r$ factorial experiments after
preliminary hypothesis testing. Australian \& New
Zealand Journal of Statistics 50, 69--79.

\rf Grieve, A.P., 1985. A Bayesian analysis of the two-period crossover
design for clinical trials. Biometrics 41, 979--990.

\rf Grieve, A.P., 1986. Corrigenda to Grieve (1985). Biometrics 42, 459.

\rf Grieve, A.P., 1987. A note on the analysis of a two-period crossover design
when the period-treatment interaction is significant. Biometrical Journal 7, 
771--776.

\rf Grizzle, J.E., 1965. The two-period change-over design and its use
in clinical trials. Biometrics 21, 467--480.

\rf Grizzle, J.E., 1974. Corrigenda to Grizzle (1965). Biometrics 30,
727.

\rf Hills, M., Armitage, P., 1979. The two-period cross-over clinical trial.
British Journal of Pharmacology 8, 7--20.

%\rf Hinkelmann, K., Kempthorne, O., 1994. Design and Analysis of
%Experiments, revised edition. John Wiley, New York.

\rf Hodges, J.L., Lehmann, E.L., 1952. The use of previous
experience in reaching statistical decisions. Annals of
Mathematical Statistics 23, 396--407.

\rf Jones, B., Kenward, M.G., 1989. Design and Analysis of Cross-Over
Trials. Chapman \& Hall, London.

%\rf Hung, H.M.J., Chi, G.Y.H., O'Neill, R.T., 1995. Efficiency
%evaluation of monotherapies in two-by-two factorial trials.
%Biometrics 51, 1483--1493.

%\rf Kabaila, P. (1995), ``The Effect of Model Selection on Confidence Regions and Prediction Regions,''
%{\it Econometric Theory}, 11, 537--549.

\rf Kabaila, P., 1998. Valid confidence intervals in regression
after variable selection. Econometric Theory 14, 463--482.

\rf Kabaila, P., 2005. On the coverage probability of confidence
intervals in regression after variable selection. Australian \&
New Zealand Journal of Statistics 47, 549--562.

\rf Kabaila, P., Giri, K., 2007a. Large sample confidence intervals in regression
utilizing prior information. La Trobe University, Department of Mathematics and
Statistics, Technical Report No. 2007--1, Jan 2007.

\rf Kabaila, P., Giri, K., 2007b. Confidence intervals in regression utilizing prior
information. {\tt arXiv:0711.3236}. Submitted for publication.

%\rf Kabaila, P., Giri, K., 2007a. Large sample confidence
%intervals in regression utilizing prior information. Technical
%Report No. 2007-1, Jan 2007, Department of Mathematics and
%Statistics, La Trobe University.

\rf Kabaila, P., Giri, K., 2008. Upper bounds on the minimum
coverage probability of confidence intervals in regression after
variable selection. To appear in
Australian \& New Zealand Journal of Statistics.

\rf Kabaila, P., Leeb, H., 2006. On the large-sample minimal
coverage probability of confidence intervals after model
selection. Journal of the American Statistical Association 101,
619--629.

\rf Kabaila, P., Tuck, J., 2008. Confidence intervals utilizing
prior information in the Behrens-Fisher problem. To appear in
Australian \& New Zealand Journal of Statistics.

%\rf Mead, R., 1988. The Design of Experiments. Cambridge
%University Press, Cambridge.

%\rf Neyman, J., 1935. Comments on `Complex experiments' by F.
%Yates. Supplement to the Journal of the Royal Statistical Society
%2, 181--247.

%\rf Ng, T-H., 1994. The impact of a preliminary test for
%interaction in a $2 \times 2$ factorial trial. Communications in
%Statistics: Theory and Methods 23, 435--450.

\rf Pratt, J.W., 1961. Length of confidence intervals. Journal of
the American Statistical Association 56, 549--657.

\rf Senn, S., 2006. Cross-over trials in {\it Statistics in Medicine}:
the first `25' years. Statistics in Medicine 25, 3430--3442.

%\rf Shaffer, J.P., 1991. Probability of directional errors with
%disordinal (qualitative) interaction. Psychometrika 56, 29--38.

%\rf Stampfer, M.J., Buring, J.E., Willett, W., Rosner, B.,
%Eberlain, K., Hennekens, C.H., 1985. The $2 \times 2$ factorial
%design: its application to a randomized trial of aspirin and
%carotene in U.S. physicians. Statistics in Medicine 4, 111---116.

%\rf Steering Committee of the Physicians' Health Study Research
%Group: Belanger, C., Buring, J.E., Cook, N., Eberlain, K.,
%Goldhaber, S.Z., Gordon, D., Hennekens, C.H. (chairman), Mayrent,
%S.L., Peto, R., Rosner, B., Stampfer, M.J., Stubblefield, F.,
%Willett, W., 1988. Preliminary report: findings from the aspirin
%component of the ongoing Physicians' Health Study. New England
%Journal of Medicine 318, 262--264.

%\rf Traxler, R.H., 1976. A snag in the history of factorial
%experiments. In: Owen, D.B. (Ed), On the History of Statistics and
%Probability, Marcel Dekker, New York, 281--295.

%\rf Tuck, J., 2006. Confidence intervals incorporating prior
%information. PhD thesis, August 2006, Department of Mathematics
%and Statistics, La Trobe University.

\end{document}